# Minimal Coarse-Grained Modelling Towards Implicit-Solvent Simulation of Generic Bolaamphiphiles

Somajit Dey[1], Jayashree Saha[2]
*Department of Physics, University of Calcutta, 92, A.P.C Road, Kolkata-700009*
Email: 1) sdphys_rs@caluniv.ac.in, 2) jsphy@caluniv.ac.in

A simple, dual-site model of bolaamphiphiles (bolaforms or bipolar amphiphiles) is developed based on an earlier single-site model of (monopolar) amphiphiles [S. Dey, J. Saha, *Phys. Rev. E* **95**, 023315 (2017)]. The model incorporates aqueous environment (both hydrophobic effect and hydration force) in its anisotropic site-site interactions, thus obviating the need to simulate solvent particles explicitly. This economy of sites and the absence of explicit solvent particles enable molecular dynamics simulations of bolaamphiphiles to achieve mesoscopic length and time-scales unattainable by any bead-spring model or explicit solvent computations. The model applies to generic bolas only, since the gain in scale can only be obtained by sacrificing the resolution of detailed molecular structure. Thanks to dual-sites, however, (as opposed to a single-site model) our model can incorporate the essential flexibility of bolas that leads to their U-conformers. The model bolas show successful self-assembly into experimentally observed nano-structures like micelles, rods, lamellae etc. and retain fluidity in very stable monolayers. Presence of membrane-spanning model bolas in bilayers of model monopolar amphiphiles increases the stability and impermeability of the lamellar phase. Model bolas are also seen to be less diffusive and to produce thicker layers compared to their monopolar counterparts. Rigid model bolas, though achiral themselves, show self-assembly into helical rods. As all these observations agree with the well-known key characteristics of archaeal lipids and synthetic bolaamphiphiles, our model promises to be effective for studies of bolas in context of biomimetics, drug-delivery and low molecular weight hydrogelators. To the best of our knowledge, no other single or dual-site, solvent-free model for bolas has been reported thus far.

## I. INTRODUCTION

*Bolaform* amphiphiles or *bola*amphiphiles (*bola* in short) are essentially bipolar molecules with their two polar moieties separated by one, two or three hydrophobic chains [1]. The generic name originates from the structural similarity these amphiphiles share with a throwing weapon, called '*bola*', consisting of weights attached to the two ends of a string. Since the identification of naturally occurring bolalipids in archaebacterial membranes, bolas have drawn wide-spread interest due to their unique properties that can aid bionanotechnology, drug and antigen delivery, gene-therapy and stimuli-responsive hydrogelation [1, 2].

Archaea are microbial extremophiles that thrive in extreme conditions like high salt concentration (halophiles), strict anaerobiosis (methanogens), high acidity (acidophiles) or high temperatures (thermophiles) as found in hot springs and under-water volcanic fields for example [3]. The presence of membrane-spanning tetraether bolalipids in their cell membranes is thought to be the key to their unusual thermal and chemical robustness [4].

Natural and model bolalipids form tightly packed yet fluid monolayers in water. This tight packing [5] and consequent low rate of lipid diffusion [6] make the membrane more impermeable to molecules and ions including $H^+$ [7], yet the fluidity keeps membrane functionality intact. Apart from the mechanical stability afforded by the monolayer organisation (absence of a preferential fracture plane), this assembly also achieves remarkable thermal stability compared to bilayers composed of standard monopolar lipids [8]. This makes liposomes formed from these monolayers amenable to thermal sterilisation without considerable leakage of their cargo [9]. Being fully saturated and lacking ester linkages, the tetraether bolalipids can also withstand oxidative stress and enzymatic activity making them ideal candidates for liposomal delivery systems [10, 11, 12, 13]. Liposomes formed from bolalipids have already shown effective transfection in gene [14, 15] and antimicrobial [16, 17] delivery. They are also seen to perform as antigen delivery systems relatively better than conventional liposomes[18, 19]. Other potential applications of bolalipid membranes related to their stability and low permeability can be found in industrial fermentation [20], functional membrane protein reconstitution [21, 22] and supported biomimetic membranes in the context of biosensors [23] or water-purifiers [24].

Well-defined bolalipids are very difficult to isolate in high quality from natural archaeal membranes (which also increases their cost) [25]. Consequently, a huge research activity is dedicated to the synthesis of chemically simple bolaamphiphiles for modelling archaeal tetraether bolas as well as for bio-inspired designs for other applications [26, 27, 28, 29, 30, 31, 32, 33, 34, 2]. The synthesised compounds range from rigid to flexible, single to multiple hydrophobic chains and acyclic to macrocyclic geometry. They may also differ in specific hydrocarbon chain modulations in order to tune membrane fluidity or in the size of hydrophilic headgroup moieties for modulating the molecular packing parameter.

Self-assembly and lyotropic behavior of these compounds as well as their mixtures with conventional monopolar amphiphiles have been extensively studied, both in bulk water and interfaces, through different experimental techniques [2, 26, 1, 35]. Bolas, depending on their packing parameter [36, 37, 2] among other things, can self-assemble into a plethora of structures including micelles, rods, tubes, ribbons/tapes, disks, vesicles and lamellar liquid crystalline phases. Some achiral bolas also have the remarkable property of forming helical nanofibers (racemic mixtures of course due to overall chiral symmetry) [38, 39, 40, 41]. By forming dense 3-dimensional networks of self-assembled nanofibers bolas can also immobilise an enormous amount of water [38, 2]. These low molecular weight hydrogelators can also be made to respond to external stimuli like pH [42], salinity [43] and temperature [44] making the gelation process switchable.

A recent review of the state-of-the-art of bolaamphiphile research and applications can be found in ref. [45]. Given this huge importance of bolas, computer simulation studies towards understanding their self-assembly, structure-function relationship and lyotropic behavior becomes highly relevant. To this end, both fine-grained and coarse-grained approaches have been taken. Fine-grained molecular dynamics simulations take into account each individual atom of the amphiphiles as well as the water molecules in the aqueous phase (all-atom models) [5, 46, 47, 48]. Consequently, an obvious drawback of this detailed approach is the highly limited system size or time-scale that can be studied. This can only be improved upon by adopting different levels of coarse-graining in the model representations of the molecules [20, 49, 50, 51]. But even then, for fully hydrated systems, water particles would take up the bulk of the computation time which again limits the scale of the bola-assemblies that can be studied. This motivates the need for implicit solvent coarse grained (ISCG) models for mesoscale simulation of amphiphiles in bulk water. Instead of positing water (i.e. the solvent) explicitly, ISCG models take into account the presence of water by recognising the hydrophobic effect as an attractive force between the amphiphiles [39, 41, 52]. Thus, the ISCG models put every computational resource at simulating the amphiphiles only, which allows for a larger system size to be studied for a longer time. It must be noted that apart from this sheer gain in scale, ISCG models also provide a broad qualitative understanding of the essential physics behind self-assembly into various lyotropic phases since it sacrifices molecular detail for generic features like hydrophilicity, hydrophobicity, packing parameter and molecular rigidity or flexibility.

We have not found in the literature any molecular dynamics (MD) simulation of bolaforms using ISCG models. However, there have been off-lattice Monte Carlo studies of a rigid bola modelled as a rigid linear chain of hydrophilic and hydrophobic beads [41, 39]. Hydrophobicity is accounted for in these works as a square-well potential attracting the hydrophobic beads together. The hydrophilic beads, on the other hand, act merely as hard spheres providing for the excluded volume interaction. Therefore, no hydration force [53] was accounted for in these simulations. This approach of using a chain of hydrophilic and hydrophobic beads can easily be extended to modelling generic bolalipids since bolalipids can always be represented as two monopolar lipids attached tail-to-tail and many such bead-spring models already exist for implicit solvent monopolar lipid simulations [54, 55, 56, 57, 58]. Although the beads interact among themselves through simple force laws, a molecule consisting of several beads is still multi-site and hence computationally expensive than a maximally coarse-grained representation with one or two sites only. A well-known example where a maximally coarse-grained (single-site) model efficiently substitutes for a linear array (multi-site) of four Lennard-Jones beads is provided by the Gay-Berne potential [59].

Apart from the monolayer-spanning *trans* configuration, most bolas can also bend themselves into *loop* or *U-shaped* conformers (whereby both headgroups of the bola remain at the same membrane-water interface). While modelling generic bolaform amphiphiles with maximal coarse-graining, this essential flexibility must be taken into account. A single-site model, however, is invariably rigid. Maximal coarse-graining thus must use atleast two single-site rigid segments attached together in a non-rigid way in a flexible dimer architecture. The present paper, in the following, reports for the first time such a model for generic bolas.

## II. MODEL

In a previous paper, we presented a single-site ISCG model for generic monopolar amphiphiles like lipids and surfactants [60]. Each amphiphile, with a hydrophilic polar head and one or more hydrophobic tails, is represented there as a soft-core directed spheroid (Fig. 1) with an anisotropic potential much like the widely-used Gay-Berne potential [59]. Thanks to a tuneable packing parameter, these model amphiphiles can spontaneously self-assemble into structures with various curvatures like micelles, rods and bilayers. The anisotropic pair-potential between the ellipsoids also features a soft short-range repulsion for certain relative orientations in order to mimic the hydration force that separates two adjacent bilayers for example. Although the length of the ellipsoids in this model is tuneable, interacting ellipsoids must be of the same length, since the minimum energy configuration of a pair of ellipsoids with different lengths does not reproduce accurate amphiphile aggregation in water. For the purpose of the present work, viz. modelling flexible bolas with maximum coarse-graining, we use two of these directed ellipsoids of equal length as subunits in a dimer architecture in the following way.

Note that any bolaform amphiphile, with two polar headgroups connected by one (two) hydrophobic chain(s), can be thought of as two monopolar amphiphiles, each having one polar head and one (two) hydrophobic tail(s), covalently bonded together at their tail ends [compare the cartoons in a) and b) of Fig. 1]. For the U-shaped configuration of the bola, these monopolar segments come side by side and become the two linear sides of the U (Fig. 1c). For the *trans* conformer, on the other hand, they remain away from each other (Fig. 1b). Motivated by this picture and our directed ellipsoid model [60] for monopolar amphiphiles as mentioned above, we model the bolas as dimers formed from two directed ellipsoids linked together at their tail-end terminals by means of a pivot (Fig. 1). When two rigid bodies have one point in common about which they can rotate freely, it is called a pivot or a spherical joint. Although the pivot allows the ellipsoids to rotate independently of each other, a V-shaped configuration leads to substantial overlap between the ellipsoids (Fig. 1c). Such an overlap, however, is opposed by their soft-core repulsion. This scheme, therefore, becomes dynamically too restrictive to model flexibility. Achieving a V-shaped configuration however remains necessary for any non-rigid dimer architecture, since that is the conformation which would correspond to the naturally occurring U-shaped or *loop* conformer in bolaamphiphiles.

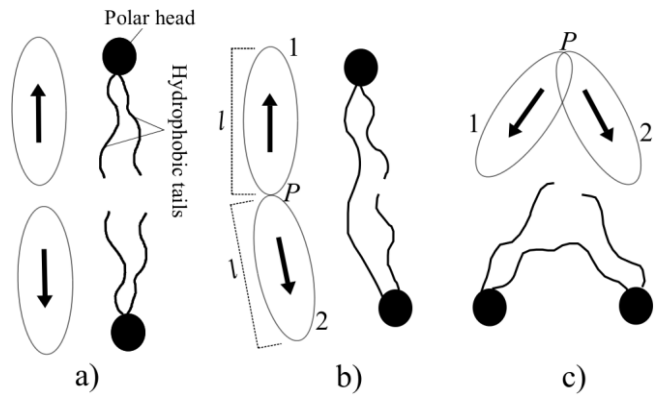

FIG. 1. a) Modelling amphiphiles with one polar headgroup as directed ellipsoids (Ref. [60]). Note the arrow is directed from tail to head. b) Modelling bipolar (bola) amphiphiles by joining two directed ellipsoids of equal length at a pivot, *P*. The bola cartoon depicts a *trans* conformer. c) *Loop* or U conformation of bola modelled as V configuration of model dimer— note the overlap at *P*.

Note that flexibility could be incorporated easily if instead of attaching the two ellipsoids at one point (viz. the pivot), we linked them through a spacer or a bond of fixed length, with no constraints on the associated bond angles [61]. But addition of a non-rigidly attached spacer like this would increase the number of degrees of freedom of the molecule which, in turn, would complicate the numerical integration of the ensuing equations of motion. Also, such a naked, one-dimensional spacer in an otherwise three-dimensional, coarse-grained molecule might call for an excluded volume interaction at the spacer region, which would then require atleast another interaction site.

The problem of overlap for V-shaped conformers in our simple pivot scheme, however, can be tackled in the following way without imposing any extra degree of freedom or requiring an additional site as above. Whenever the two directed ellipsoids in a bola interact with each other, let us replace each of them with a shorter directed spheroid attached rigidly to a linear spacer along its axis at the tail-end, as illustrated in Fig. 2. In other words, for the *intra*-bola interaction, each of the original spheroidal subunits (Fig. 1) is being resolved into an interacting spheroidal core at the top and a non-interacting region at the bottom which is attached to the pivot (Fig. 2a). The new spheroidal core may be identical to its original counterpart in all the interaction parameters [ $\varepsilon_0$, $v_0$, $v_1$, $v_2$, $v_3$, $\varepsilon_e$ ] except in its length. To retain the total length of the whole molecule, however, the length of each of these new ellipsoids must be shorter than its original counterpart just by the length of its spacer ( $l_{\text{spacer}}$ ) (Fig. 2a). Incorporation of these rigid spacers eliminates overlap between the interacting cores in the V-shaped conformer thus helping flexibility (Fig. 2b). Yet it does not add any new degree of freedom to the model molecule, since the spacers remain rigidly attached to the spheroids along their axes of symmetry. Note that the spacers are considered only when we deal with the mutual interaction between the two constituent subunits within the same bola (i.e. *intra* dimer interaction), but not while subunits from different bolas (i.e. *inter* dimer interaction) interact. Because of this, the other amphiphiles do not *see* the spacers within a bola molecule. In other words, since only the larger ellipsoids (Fig. 2) are used instead of the spacer and smaller spheroid combination for inter-amphiphile interactions, there remains no need to provide an additional excluded volume interaction at the spacers in order to prevent the other amphiphiles from interfering in that region.

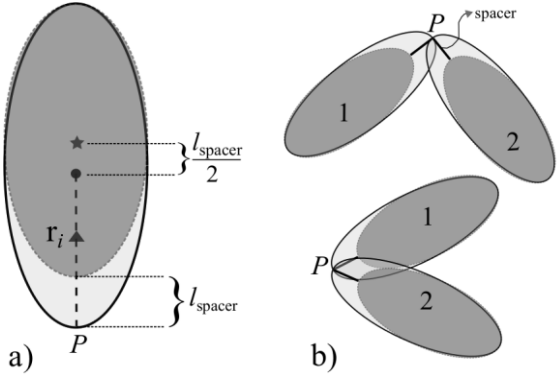

FIG. 2. Construction for intra-dimer case: a) Detailed resolution of any subunit ($i$ = 1, 2). The large ellipsoid (enclosed in solid line) is resolved into a smaller ellipsoid (darker shade and enclosed in dotted line) at the top and a non-interacting region (lighter shade) near the pivot, $P$. The circle denotes the centre of the large ellipse and the star denotes that of the smaller ellipse. Note that the large and the small ellipse differ only in height but not width. b) In V-configurations, only the non-interacting bottom regions (lighter shade) overlap but the interacting top ellipsoids (darker shade) do not.

To summarise, our model bola consists of two directed spheroids each of which individually models a monopolar amphiphile as described in Ref. [60]. The two spheroids are of equal length and are connected at their tail ends by a pivot (Fig. 1). Each such bola can interact with other bolas, or even monopolar amphiphiles modelled as directed spheroids, through pair-interactions between their respective spheroids according to the parametric anisotropic potential of Ref. [60]. While considering the interaction between the two constituent spheroids within the same bola dimer (*intra* bola interaction), however, the spheroids are shortened length-wise at their tail ends so that each of them is now connected by a spacer at the pivot (Fig. 2). Lengths of the two spacers (and consequently of the shortened spheroids) are equal. This spacer length, however, can be seen as a tuneable parameter of the bola model since it affects how close the two spheroids can come to each other (i.e. how tight an angle they make at the pivot) in the *loop* conformation.

To aid molecular dynamics studies of this flexible dimer model, a discussion on the equations of motion follows in the next section. It may be remarked that only a very involved scheme exists in the literature for the molecular dynamics of any serial chain multibody composed of rigid subunits hinged together [62]. Although our flexible dimer model falls under the purview of that scheme, the equations of motion proposed originally in the following are intended to be more convenient and simpler. We also present a numerical integration scheme for our equations of motion in Appendix A.

<u>Symmetric and asymmetric bolas:</u> Note that although all interacting spheroids must be of the same length in our proposed model, different interacting pairs may have different packing parameters (governed by the parameter $\varepsilon_e$ [60]). For example, different packing parameters for the constituent spheroids of a bola dimer models for differing sizes of the two headgroups in a bolaamphiphile. Our dimer model, therefore, can model both *symmetric* bolas (by having all the parameters [$\varepsilon_0$, $v_0$, $v_1$, $v_2$, $v_3$, $\varepsilon_e$] equal for both the constituent spheroids) and *asymmetric* bolas (by having some parameters differ). Later in this paper (Sec. IV), however, we report simulations of symmetric bolas only.

<u>Note on rigid bola:</u> The dimer model for flexible bolas as discussed above can be conveniently used to model rigid bolas just by removing the flexibility at the pivot. For linear rigid bolas, the two subunits always remain anti-parallel to each other and the whole molecule behaves as a linear rigid body with only 5 degrees of freedom. Since the subunits never overlap for such a linear dimer, we no more need to implement spacers to limit the intra-dimer interactions in this case.

## III. EQUATIONS OF MOTION

Since the directed spheroids in our dimer bola-model are symmetric about their long axes [60], they can be dynamically represented as linear rigid rotors characterised by a mass and moment of inertia. The position vector of any point fixed on the long axis and the unit vector along the directed long axis suffice to describe the position and orientation of each spheroid individually. Monopolar amphiphiles modelled as such directed spheroids [60] follow the simple unconstrained dynamics of linear rigid bodies as detailed in Ref. [63]. Motion of the two spheroids in our model bola dimer, however, is constrained due to the presence of the pivot, $P$. Unlike unconstrained rigid body motion, this leads to coupling between the translational and rotational dynamics of the two subunits in the dimer as will be seen below.

The state of the dimer can be completely specified by the set $(\mathbf{r}, \dot{\mathbf{r}}, \mathbf{r}_1, \mathbf{r}_2, \boldsymbol{\omega}_1, \boldsymbol{\omega}_2)$, where $\mathbf{r}$ and $\dot{\mathbf{r}}$ denote the position and velocity of the pivot, $P$, and $\mathbf{r}_1$ and $\mathbf{r}_2$ denote the position vectors of the respective centres of mass of the spheroidal subunits *relative* to $P$ (Fig. 2a). $\boldsymbol{\omega}_1$ and $\boldsymbol{\omega}_2$ denote the angular velocities of the subunits considered as linear rotors rotating about $P$. Hence, $\boldsymbol{\omega}_i$ must always be perpendicular to $\mathbf{r}_i$. The equations of motion for this set are:

$$M\ddot{\mathbf{r}} = \mathbf{F} + \sum_{i=1}^{2} m_i \omega_i^2 \mathbf{r}_i - \sum_{i=1}^{2} m_i \dot{\boldsymbol{\omega}}_i \times \mathbf{r}_i \quad (1)$$

and

$$I_i \dot{\boldsymbol{\omega}}_i = \boldsymbol{\Gamma}_i + \mathbf{r}_i \times (-m_i \ddot{\mathbf{r}}). \quad (2)$$

Above, $\mathbf{F}$ is the total force on the dimer, $M$ is the total mass and $m_i$ and $I_i$ denote the mass and moment of inertia (about $P$) of the two individual subunits respectively for $i=1$ and $2$. $\boldsymbol{\Gamma}_i$ denotes the total inertial torque experienced by the $i$-th subunit about $P$. Note that $I_i$ can be obtained from $I_i^C$, the moment of inertia of the $i$-th subunit about its centre of mass, as

$$I_i = I_i^C + m_i |\mathbf{r}_i|^2. \quad (3)$$

Note that while the total force on the bola-dimer is the vector sum of the forces $\mathbf{F}_i$ experienced by each of the two subunits, i.e.

$$\mathbf{F} = \mathbf{F}_1 + \mathbf{F}_2, \quad (4)$$

$\mathbf{F}_i$ is composed of forces experienced by the $i$-th subunit due to both inter- and intra-bola interactions, viz.

$$\mathbf{F}_i = \mathbf{F}_i^{\text{inter}} + \mathbf{F}_i^{\text{intra}}. \quad (5)$$

Most importantly, however, these two forces differ in their point of application. $\mathbf{F}_i^{\text{inter}}$ applies on the centre of mass of the $i$-th spheroidal subunit (i.e. the dot in Fig. 2a), while $\mathbf{F}_i^{\text{intra}}$ applies on the centre of mass of the shorter spheroid that the former is resolved into during intra-bola considerations (i.e. the star in Fig. 2a). $\boldsymbol{\Gamma}_i$ is thus computed as:

$$\boldsymbol{\Gamma}_i = \mathbf{T}_i^{\text{inter}} + \mathbf{T}_i^{\text{intra}} + \mathbf{r}_i \times \mathbf{F}_i^{\text{inter}} + (|\mathbf{r}_i| + \frac{l_{\text{spacer}}}{2})\hat{\mathbf{r}}_i \times \mathbf{F}_i^{\text{intra}} \quad (6)$$

where $l_{\text{spacer}}$ denotes the spacer length and $\hat{\mathbf{r}}_i$ is the unit vector along $\mathbf{r}_i$. The $\mathbf{T}_i$'s above denote the torque generated due to the orientation dependence of the anisotropic interaction potential. Expressions for the $\mathbf{F}_i$'s and $\mathbf{T}_i$'s can be found in Appendix of Ref. [60]. It maybe remarked here that since $\sum_{i=1}^{2} \mathbf{F}_i^{\text{intra}} = 0$, the intra-bola forces do not contribute to $\mathbf{F}$.

Numerical integration of a set of equations of motion requires computation of the accelerations for any given instantaneous state. In order to solve for the linear and angular accelerations from the instantaneous coordinates and velocities, however, the equations of motion (1) and (2) must be decoupled. Eliminating $\dot{\boldsymbol{\omega}}_i$ in Eq. (1) by using Eq. (2) and simplifying, we get

$$M\ddot{\mathbf{r}} + \sum_{i=1}^{2}\frac{m_i}{I_i}\mathbf{r}_i \times (\mathbf{r}_i \times m_i\ddot{\mathbf{r}}) = \mathbf{F} + \sum_{i=1}^{2}m_i\left(\omega_i^2\mathbf{r}_i - \frac{1}{I_i}(\boldsymbol{\Gamma}_i \times \mathbf{r}_i)\right) \quad (7)$$

Note that in this equation of motion $\ddot{\mathbf{r}}$ does not depend on $\dot{\boldsymbol{\omega}}_i$ anymore. Since $\mathbf{r}_i \times (\mathbf{r}_i \times m_i\ddot{\mathbf{r}}) = m_i(\mathbf{r}_i \cdot \ddot{\mathbf{r}})\mathbf{r}_i - m_i r_i^2 \ddot{\mathbf{r}}$, Eq. (7) reads in the operator format as

$$\sum_{i=1}^{2}\left(m_i\frac{I_i^C}{I_i} + \frac{m_i^2}{I_i}\mathbf{r}_i\mathbf{r}_i \cdot \right)\ddot{\mathbf{r}} = \mathbf{F} + \sum_{i=1}^{2}m_i\left(\omega_i^2\mathbf{r}_i - \frac{1}{I_i}(\boldsymbol{\Gamma}_i \times \mathbf{r}_i)\right)$$

or (8)

$$A\ddot{\mathbf{r}} = \mathbf{B}$$

In matrix representation, A becomes

$$A \triangleq \sum_{i=1}^{2}\left(m_i\frac{I_i^C}{I_i}\mathbf{I}_3 + \frac{m_i^2}{I_i}\begin{pmatrix}r_i^x \\ r_i^y \\ r_i^z\end{pmatrix}\begin{pmatrix}r_i^x & r_i^y & r_i^z\end{pmatrix}\right)$$

$$= \begin{pmatrix} z + \sum_{i=1}^{2}\frac{m_i^2}{I_i}(r_i^x)^2 & \sum_{i=1}^{2}\frac{m_i^2}{I_i}r_i^x r_i^y & \sum_{i=1}^{2}\frac{m_i^2}{I_i}r_i^x r_i^z \\ \sum_{i=1}^{2}\frac{m_i^2}{I_i}r_i^x r_i^y & z + \sum_{i=1}^{2}\frac{m_i^2}{I_i}(r_i^y)^2 & \sum_{i=1}^{2}\frac{m_i^2}{I_i}r_i^y r_i^z \\ \sum_{i=1}^{2}\frac{m_i^2}{I_i}r_i^x r_i^z & \sum_{i=1}^{2}\frac{m_i^2}{I_i}r_i^y r_i^z & z + \sum_{i=1}^{2}\frac{m_i^2}{I_i}(r_i^z)^2 \end{pmatrix} \quad (9)$$

where $z = \sum_{i=1}^{2} m_i \frac{I_i^C}{I_i}$. Note that A is symmetric which eases the process of both computing it and inverting it. Multiplying the vector on the right hand side of Eq. (8), i.e. $\mathbf{B}$, with $A^{-1}$ solves for $\ddot{\mathbf{r}}$. Using this value of $\ddot{\mathbf{r}}$, one can then proceed to determine $\dot{\boldsymbol{\omega}}_i$ from Eq. (2).

To summarise the above discussion and for ease of access, we reformulate the equations of motion for our bola-dimer as follows:

$$\ddot{\mathbf{r}} = A^{-1}\mathbf{B} \quad (10)$$

and

$$\dot{\boldsymbol{\omega}}_i = (\boldsymbol{\Gamma}_i + \ddot{\mathbf{r}} \times m_i\mathbf{r}_i)/I_i. \quad (11)$$

For the sake of completeness, we also note that the kinetic energy of the bola-dimer is given by

$$\frac{1}{2}M\dot{\mathbf{r}}^2 + \frac{1}{2}\sum_{i=1}^{2}I_i\boldsymbol{\omega}_i^2 + \dot{\mathbf{r}}\cdot\sum_{i=1}^{2}m_i\boldsymbol{\omega}_i \times \mathbf{r}_i. \quad (12)$$

Numerical integration of the above equations of motion is discussed in Appendix A. To perform constant temperature molecular dynamics on a system of bolas, it is also necessary to couple the system to a thermostat. This can be done either by direct rescaling of the velocities, by introducing stochastic collisions or by extending the system with an extra degree of freedom representing the heat-bath [64]. Appendix B discusses an extended system scheme that is inspired by the Nosé-Hoover thermostat [65, 66].

## IV. MOLECULAR DYNAMICS (MD) SIMULATIONS

In the following, we report the results of simulations with our implicit-solvent dimer model for flexible and rigid bolas. Note, however, that our primary goal in these initial simulations have been just to see if our maximally coarse-grained model, with only implicit hydrophobic and hydration forces, can successfully reproduce some of the key features of generic bola-assemblies as established through experiments and much more elaborate multi-site computer models with explicit solvents. More specifically, we focussed only on the general self-assembly behavior from random gas phases, along with a few key physical properties of the lamellar phase like stability, packing, ordering, layer thickness and fluidity via self-diffusion. In view of this, an extensive exploration of the parameter-space has not been done in this early study but might be undertaken in the future.

Mainly three systems were studied and compared: *a*. a system composed of identical bolas, *b*. a mixture of bolas and monopolar amphiphiles and *c*. a system of identical monopolar amphiphiles. For simplicity, the monopolar amphiphiles used in our simulations would henceforth be referred to as 'lipids' and the bipolar or bolaamphiphiles as 'bola's. Bolas in systems *a* and *b* were modelled as flexible dimers as described in Sec. II. The lipids in systems *b* and *c* were modelled as directed spheroids following Ref. [60]. For the sake of comparison, the same parameters were used for the pair-interactions between the spheroids in all of the above systems. Note that this makes all the bolas symmetrical. That choice of parameters was adopted which was shown to generate self-assembled lamellar phases in Ref. [60]: $\nu_0 = 1$, $\nu_1 = 0.8$, $\nu_2 = 4$, $\nu_3 = 3$, $\varepsilon_e = 0.052$, $\sigma_e = 3$, $range = 3$. Note that for the above choice of parameters, the length of each bola is exactly the double that of a lipid. The spacer length for all intra-dimer cases (Fig. 2a) was chosen as $l_{spacer} = 0.5$. For the intra-dimer interactions, the smaller ellipsoids thus have length, $\sigma_e = 3 - l_{spacer} = 2.5$. In order to work in reduced units, we also chose $\sigma_0 = 1$, $\varepsilon_0 = 1$ and the mass of each ellipsoid ($m_i$) = 1. To ensure that the time-scales for translational and rotational dynamics are comparable to each other for all the ellipsoids, the moment of inertia about their centres of mass ($I_i^C$) was chosen to be 4. [In addition to the systems *a-c*, a system of rigid linear bolas was also studied using the above specifications and will be reported later in this section].

Both constant energy (NVE) and constant temperature (NVT) molecular dynamics simulations were performed. Constant pressure molecular dynamics could not be applied due to the absence of an explicit aqueous phase for pressure coupling. For flexible bolas, we used the leap-frog schemes discussed in Appendix A and B. For lipids and rigid linear bolas, the leap-frog schemes presented in Ref. [63] were followed. The integration time-step, $h$, was chosen to be 0.0035 for all runs. For NVT runs, $\tau$ was taken as 0.07 for each of the individual temperature controls (Appendix B). Given the novelty of the integrators used, we demonstrate their stability and temperature control in Fig. 3.

Each of the systems under study was kept confined in a cubic simulation box with periodic boundary conditions. The velocity initialisations made sure that the total momentum and total angular momentum were both null. All algorithms were implemented in FORTRAN 95 with double precision computation for each floating-point operation.

The systems *a-c* were initially setup in lamellar configurations as follows. The simulation box was trisected by two horizontal square grids half a box-length apart with each grid containing 400 cells. Each grid-point was now covered with either a bola or a set of two

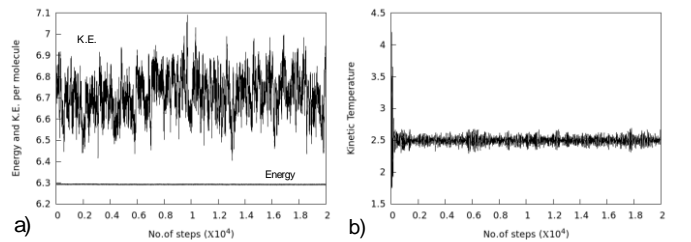

FIG. 3 a) Plot of energy and kinetic energy for a typical NVE run. Energy plot is shifted vertically to shorten the diagram height. Note that the fluctuation in energy is insignificant compared to that in kinetic energy. Also no significant energy drift is present. b) Plot of kinetic temperature for an NVT run with $\Theta = 2.5$.

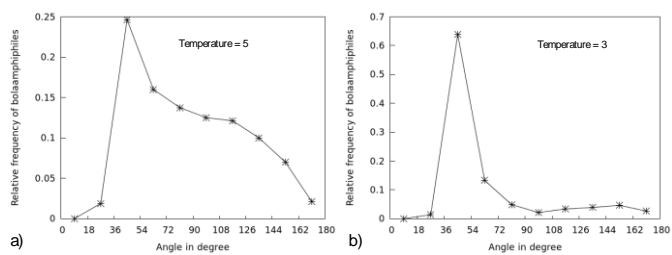

FIG. 4 Histogram plotting relative frequency distribution of the angle between the subunits in a bola-dimer. Small angles denote *loop* and wide angles imply *trans* conformer. a) For $\Theta = 5$ b) For $\Theta = 3$.

lipids directed away from each other. All ellipsoids were kept normal to the grid-plane. Bolas were positioned at the grid-points on their pivots while the lipids were positioned on their tail-ends. The pure bola system, *a*, thus contains 800 bolas in two monolayers, while the pure lipid system, *c*, contains 1600 lipids in two bilayers. For the mixed system, *b*, 133 bolas were distributed randomly on each grid while the remaining grid-points were filled with lipids. This system thus contains 266 bolas and 1068 lipids. The distance between the neighbouring grid-points, viz. the cell-size, was chosen to be 1.2. With this choice, all the initial configurations remained free of overlap as desired. In addition, the top and bottom layers remained out of range of each other's repulsion. What more, the resulting density ensured that the pressure, computed using the virial [64], remained near $0 \pm 0.5$ for the temperatures reported here. Such zero pressure implies that the amphiphiles could freely arrange themselves into preferred configurations despite the constancy of box-size.

In order to study self-assembly in systems *a* and *b*, we generated randomised isotropic gas configurations for both the systems by melting their pre-built lamellar configurations at a high temperature ($\Theta = 5$). Note that this way of producing random configurations automatically ensures the required non-overlap at a reasonable density for implicit solvent systems.

For *a*, the bolas remained randomly distributed throughout the simulation box at $\Theta = 5$ and the angle between the two subunits in a dimer showed a broad frequency distribution (Fig. 4a). Upon cooling at $\Theta = 4$, system *a* showed discernible self-assembly, although no clear structure emerged. At $\Theta = 3.5$, however, clear rod-like fibres appeared (Fig. 5b Left) with a predominance of the *loop* or V-conformation (Fig. 4b). As expected, the fibres got more pronounced as the system was cooled further. Rods and micelles (Fig. 5a) composed of V-shaped bolas were also found during some early test runs from an FCC initialisation at a lesser number density (= 0.05). It stands to reason, therefore, that self-assembly occurs as the following. Isolated bolas choose a *loop* conformation over a *trans* conformation in order to minimise the hydrophobic area exposed to the aqueous solvent. (Such bending prior to self-assembly is also suggested by the explicit solvent simulations of Ref. [20]). These conformers then self-assemble into micelles which, in turn, fuse to form rods and consequently fibres. Note that for the pure bola system (*a*) self-assembly into a lamellar phase (monolayer or bilayer) was not observed. This is in agreement with the experimental fact that short-chain or single chain, flexible, fully-hydrated bolas self-assemble into micelles or thread-like micelles (viz. rods and fibres) [1, 24, 67, 68].

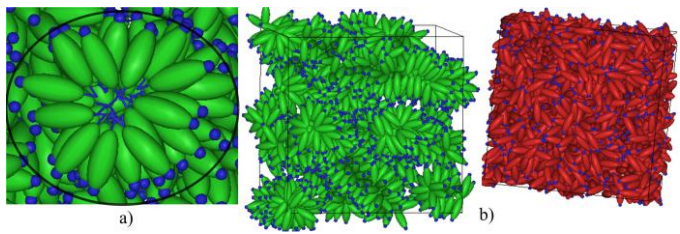

FIG. 5 (Colour online) a) Cross-section of a micelle formed by self-assembly of bola-dimers in system *a*. Blue spheres mark the polar ends of the ellipsoids. Spacers (blue rods) are shown for clarity. Note that the dimers adopted *loop* conformation. b) Comparison of thermal stability at $\Theta = 3.5$— Left: Structure visible in system *a*; Right: Random isotropic gas phase in system *c*. Green denotes bola, red denotes lipid.

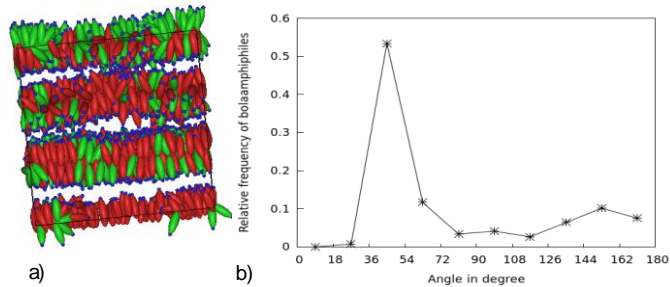

FIG. 6 (Colour online) a) Self-assembly into lamellar phase in system *b* at $\Theta = 2.5$. b) Histogram for bending angle distribution for the configuration in a).

Upon cooling from the randomised configuration, system *b* however showed a different self-assembly behaviour. Thanks to the majority presence of bilayer forming lipids, system *b* self-assembled into lamellar phases composed of a mixture of lipids, *trans* bolas and V-shaped bolas (Fig. 6a). The presence of lipids is also reasoned to stabilise *trans* bolas, as the relative frequency of *trans* bolas was found to be greater than that in system *a* (Fig. 6b). The onset of structure is also distinctly different for *b* as compared to *a*. Whereas structure was discernible in *a* as early as $\Theta = 4$, structure appeared in *b* only upon further cooling. With complete absence of bolas, system *c* was even less stable. This shows that bolas add to the thermal stability of the assemblies (Fig. 5b).

Simulations of the membrane-lamellar phases were initiated directly from the pre-built (grid-based) lamellar phases, but results were taken only after adequate thermalisation ($\approx 10^5$ MD steps). Final configuration of one run was used as the initial configuration of the subsequent run wherever possible. Unlike Ref. [20], however, no deliberate equilibration between the *trans* and *loop* conformers was attempted. At $\Theta = 2.5$, system *a* showed stable monolayer with pores (Fig. 7a). Note that although the monolayer is predominantly made of *trans* bolas, the pore edges are formed by V-shaped bolas.

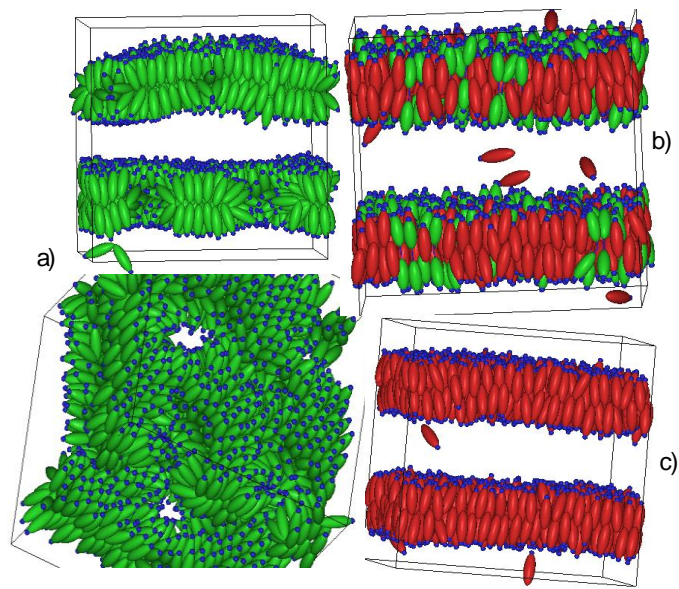

FIG. 7 (Colour online) Configurations at $\Theta = 2.5$. a) Monolayer in system *a* showing pores. No pore formation in b) system *b* and c) system *c*. Notice interdigitation of red lipids in b) and c).

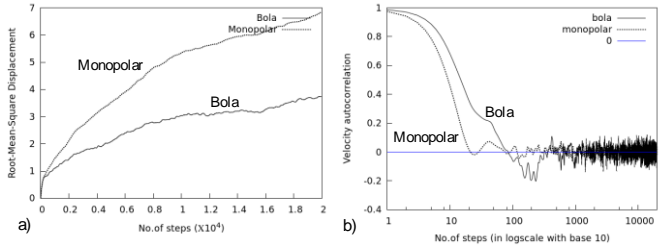

FIG. 8 Fluidity in lamellar phase at $\Theta = 2.5$. a) R.M.S displacement b) Velocity autocorrelation function in the layer-plane.

At the same temperature, system *b* and *c* however showed stable layers with no pores (Fig. 7b and c). Since area of the whole layer (including the pores, if any) is the same for both *a* and *c*, the above observation implies that bolas pack more tightly than lipids. This explains the increased impermeability of bola monolayers as compared to lipid bilayers.

Orientational (uniaxial) order parameter for lipids in a bilayer is given as $S = \frac{1}{2}\langle 3(\hat{\mathbf{u}}_i \cdot \hat{\mathbf{n}})^2 - 1 \rangle_i$ where $\hat{\mathbf{u}}_i$ denotes the unit vector along the axis of the $i$-th lipid and $\hat{\mathbf{n}}$ denotes the average bilayer normal [54]. Orientational order of our bola-dimers in lamellar phase was measured by the uniaxial order parameter of their lipid-like spheroidal subunits computed as above. Lipids and bolas were found to have similar order: $S \approx 0.898$ for lipids and $S \approx 0.880$ for bolas. Thickness of a layer was estimated as $\delta = 2\sigma_{\hat{\mathbf{n}}}$ where $\sigma_{\hat{\mathbf{n}}}$ denotes the root-mean-square normal distance of the head-ends of the directed spheroids from the mean layer plane. We found $\delta_{\text{lipid}} \approx 5.515$ and $\delta_{\text{bola}} \approx 6.035$. Hence, a bola monolayer is thicker than a corresponding lipid bilayer. Given the similarity of orientational order of bolas and lipids, this difference in thickness may be explained by the interdigitation of lipids from the two leaflets of the bilayer (Fig. 7c).

We studied fluid lamellar phases only. The fluid-gel transition at low temperatures was not explored. Fluidity in a layer is demonstrated in Fig. 8a by the increasing root-mean-square displacements of the pivots for bolas and the centres of mass for lipids. Liquidity also becomes apparent with the rapid decay of the corresponding velocity autocorrelation functions in the layer plane (Fig. 8b). The self-diffusion constant, $D$ was measured from an NVE run using the Einstein relation [64, 69]. For system *b*, at $\Theta = 2.5$, we obtained $D_{\text{lipid}} \approx 0.533$ and $D_{\text{bola}} \approx 0.033$. It may be remarked that such an order-of-magnitude difference in the diffusion coefficients of tetraether bolalipids and monopolar diether lipids is also observed in the all-atom, explicit solvent MD simulations of Ref. [47].

Rigid linear bola:

From early experiments and computer simulations [38, 39, 40, 41], dumb-bell shaped rigid bolas with bulky headgroups are known to form helical rods. The headgroups of each subunit in our bola-dimer can be made bulkier by increasing the parameter $\varepsilon_e$, while keeping all the other parameters the same. As shown in Ref. [60], for a choice of $\varepsilon_e \geq 0.11$, the lipid like ellipsoids self-assembled into curved phases like rods (Fig. 9a) and micelles (Fig. 9b), characterised by lower packing parameters or bulkier headgroups. Dumb-bell shaped rigid bolas in our dimer model, thus, could be represented with $\varepsilon_e \geq 0.12$ for both the subunits (other parameters being equal to what was reported before in this section). We report NVT MD with 500 of these rigid linear dimers at a number density, 0.05. Mass of each rigid dimer was taken as unity and *P* was taken as the centre of mass. $I_i^C$ was taken as 2 and the integration time step, $h = 0.0035$. From a randomised isotropic configuration, generated by melting an FCC lattice, these rigid dimers indeed self-assembled into helical rods for temperature $\Theta \leq 2.5$ (Fig. 9c).

## V. CONCLUSION

The importance of implicit solvent coarse-grained (ISCG) models for simulating complex systems at large enough scales cannot be overestimated. Such low resolution modelling has been successfully done in polymer physics and biological systems [70]. Given the continued interest in bolaamphiphile research (Sec. I), we proposed an ISCG dimer model for bolas in this paper.

With only two interaction sites (for both inter- and intra- dimer interactions), our model is kept minimalistic in its pair-potential and molecular architecture which simplifies the force computations and equations of motion respectively. This should therefore benefit large scale molecular dynamics (MD) of generic bolas. Despite its simplicity and solvent-free nature, our model successfully generated many well established features of bolaamphiphile systems including self-assembly behavior, thermal stability, greater layer thickness, lower fluidity and higher packing order. As discussed in the paper, our simple dimer can be used to model flexible or rigid bolas on one hand, and symmetric or asymmetric bolas on the other. The packing parameter of each subunit of the model bola dimer is tuneable which also adds to its versatility.

Mixture of bolas with monopolar amphiphiles can be studied with our model as shown in the paper. Mixtures of different species of bolas may also be studied by choosing the parameters for inter-species interactions differently from those used for the intra-species case. A drawback of our model, however, is the stipulation of identical lengths for all interacting spheroids. Due to this, our model may not be used, for example, to study domain formation or phase separation driven by hydrophobic mismatch in mixtures of bipolar and monopolar lipids [71].

We are currently working on the application of bolaamphiphilies in carrier design for drug and gene delivery using this model. Though starting with flexible symmetric bolas for simplicity, our work may probe into asymmetric bolas in future, if required.

### ACKNOWLEDGEMENT

S.D. is funded by the Council of Scientific & Industrial Research (CSIR), India, through a Senior Research Fellowship [File No. 09/028(0960)/2015-EMR-I].

### APPENDIX A:
### Numerical integration of the equations of motion (10) and (11)

Note that if $\ddot{\mathbf{r}}$ in Eq. (10) did not depend on $\boldsymbol{\omega}_i$ (through $\mathbf{B}$), but only on $\mathbf{r}$ and $\hat{\mathbf{r}}_i$, then the evolution of $\mathbf{r}$ and $\dot{\mathbf{r}}$ could be computed through a simple leap-frog or velocity-Verlet algorithm [64, 69]. A similar simple scheme [63] would apply for the evolution of $\hat{\mathbf{r}}_i$ and $\boldsymbol{\omega}_i$ too. Even if $\ddot{\mathbf{r}}$ and thereby $\dot{\boldsymbol{\omega}}_i$ depended on $\boldsymbol{\omega}_i$ only linearly, a simple time-reversible leap-frog scheme could still be devised for the evolution of $\hat{\mathbf{r}}_i$ and $\boldsymbol{\omega}_i$, albeit involving velocity-independent matrices, as done in case of the Nosé-Hoover equations of motion where angular acceleration depends on angular velocity linearly (see Appendix C of Ref. [63]). The non-linear dependence of $\ddot{\mathbf{r}}$ and consequently of $\dot{\boldsymbol{\omega}}_i$ on $\boldsymbol{\omega}_i$, however, precludes a non-iterative symplectic time-reversible scheme for the numerical integration of the equations of motion (10) and (11). It may be remarked that such non-linear dependence of accelerations on velocities is a characteristic of internal variable molecular models in general [62].

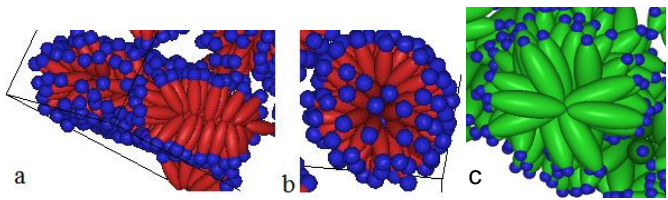

FIG. 9 (Colour online) a) Rods ($\varepsilon_e = 0.11$) and b) micelles ($\varepsilon_e = 0.15$) formed from lipid-like ellipsoids [60]. c) Helical rods formed by rigid linear symmetric bola dimers with $\varepsilon_e = 0.12$.

The standard time-reversible velocity-Verlet is a third-order integrator because it propagates the on-step state variables with error of the order $h^3$, where $h$ denotes the integration time-step [64, 69]. In the following, we describe a similar third-order algorithm to integrate the equations of motion (10) and (11). Starting from an $O(h^3)$-accurate knowledge of $\mathbf{r}(t), \dot{\mathbf{r}}(t), \ddot{\mathbf{r}}(t)$ and $\hat{\mathbf{r}}_i(t), \boldsymbol{\omega}_i(t), \dot{\boldsymbol{\omega}}_i(t)$ at time $t$, our goal is thus to find their $O(h^3)$-accurate estimates at time $t+h$. This is achieved through the following steps. [Truncation errors are shown in square brackets to keep track of the accuracy of the estimates].

Velocity-Verlet scheme:

*Step* 1: Update velocities to mid-step as:
$$\dot{\mathbf{r}}(t+\frac{h}{2}) = \dot{\mathbf{r}}(t) + \frac{h}{2}\ddot{\mathbf{r}}(t) \qquad [O(h^2)] \qquad (A1)$$
$$\boldsymbol{\omega}_i(t+\frac{h}{2}) = \boldsymbol{\omega}_i(t) + \frac{h}{2}\dot{\boldsymbol{\omega}}_i(t) \qquad [O(h^2)] \qquad (A2)$$

*Step* 2: Propagate coordinates and orientations by a full step:
$$\mathbf{r}(t+h) = \mathbf{r}(t) + h\dot{\mathbf{r}}(t+\frac{h}{2}) \qquad [O(h^3)] \qquad (A3)$$
$$\hat{\mathbf{r}}_i(t+h) = \frac{\hat{\mathbf{r}}_i(t)\left(1 - \frac{\theta^2}{4}\right) + \boldsymbol{\theta}}{\left(1 + \frac{\theta^2}{4}\right)} \qquad [O(h^3)] \qquad (A4)$$

where $\boldsymbol{\theta} = h\boldsymbol{\omega}_i(t+h/2) \times \hat{\mathbf{r}}_i(t)$ and $\theta = |\boldsymbol{\theta}|$ [63].

*Step* 3: Compute the forces (i.e. $\mathbf{F}_i(t+h)$) and torques (i.e. $\mathbf{T}_i(t+h)$) from the on-step positions as updated in step 2. Then find $\mathbf{F}(t+h)$ using Eq. (4) and $\boldsymbol{\Gamma}_i(t+h)$ using Eq. (6). Also compute $\mathrm{A}^{-1}(t+h)$ using Eq. (9). [Accuracy of each estimate is $O(h^3)$].

*Step* 4: Approximate $\ddot{\mathbf{r}}(t+h)$ as $\ddot{\mathbf{r}}(t)$, viz.
$$\ddot{\mathbf{r}}(t+h) = \ddot{\mathbf{r}}(t) \qquad [O(h)] \qquad (A5)$$
and compute an estimate of $\dot{\boldsymbol{\omega}}_i(t+h)$ using Eq. (11) [with $O(h)$ accuracy]. (Note that this way of estimating the angular acceleration keeps it perpendicular to the linear rotor axis i.e. $\hat{\mathbf{r}}_i$).

*Step* 5: Estimate $\boldsymbol{\omega}_i(t+h)$ as
$$\boldsymbol{\omega}_i(t+h) = \boldsymbol{\omega}_i(t+\frac{h}{2}) + \frac{h}{2}\dot{\boldsymbol{\omega}}_i(t+h) \qquad [O(h^2)] \qquad (A6)$$
(Note that $\boldsymbol{\omega}_i(t+h)$ remains orthogonal to $\hat{\mathbf{r}}_i(t+h)$ as expected).

*Step* 6: Compute $\ddot{\mathbf{r}}(t+h)$ (Eq. (10)) and therefrom determine $\dot{\boldsymbol{\omega}}_i(t+h)$ (Eq. (11)) [with $O(h^2)$ order of accuracy].

*Step* 7: Repeat step 5 using this [$O(h^2)$] estimate of $\dot{\boldsymbol{\omega}}_i(t+h)$ from step 6. [Note that Eq. (A6) now becomes $O(h^3)$ accurate].

*Step* 8: Repeat step 6, using the current estimate of $\boldsymbol{\omega}_i(t+h)$, to get an $O(h^3)$ accurate estimate of $\ddot{\mathbf{r}}(t+h)$ and $\dot{\boldsymbol{\omega}}_i(t+h)$. Save the $O(h^3)$ estimate of $\boldsymbol{\omega}_i^2(t+h)$ that would have to be computed in order to determine $\ddot{\mathbf{r}}(t+h)$.

*Step* 9: Update velocities from mid-step to on-step as:
$$\dot{\mathbf{r}}(t+h) = \dot{\mathbf{r}}(t+\frac{h}{2}) + \frac{h}{2}\ddot{\mathbf{r}}(t+h) \qquad [O(h^3)] \qquad (A7)$$
$$\boldsymbol{\omega}_i(t+h) = \boldsymbol{\omega}_i(t+\frac{h}{2}) + \frac{h}{2}\dot{\boldsymbol{\omega}}_i(t+h) \qquad [O(h^3)] \qquad (A8)$$

This completes the evolution of the dynamical state for one integration time-step. The kinetic energy at time $t+h$ can be computed now from Eq. (12) using the $\boldsymbol{\omega}_i^2(t+h)$ which had been saved during step 8. Note that to initiate the above algorithm, we must compute $\ddot{\mathbf{r}}(t=0)$ and $\dot{\boldsymbol{\omega}}_i(t=0)$ from the initial positions and velocities using Eq. (10) and (11).

Although this integration scheme is not time-reversible, the forward trajectory differs locally from the time-reversed trajectory only on the order of $O(h^4)$, which is smaller than the third-order local truncation error of the algorithm itself. It may be remarked however that even if our integrator was reversible, the finite precision of practical computation would nevertheless render it irreversible.

Leap-Frog scheme:

A leap-frog version of this velocity-Verlet algorithm can easily be arrived at by rearranging the steps 1-9 and hence would not be elaborated for brevity. It may be remarked, however, that instead of separate execution of steps 9 and 1 (of the next cycle), leap-frog executes the final velocity update by a full step from one mid-step to the next [64, 69]. Thus instead of Eq. (A7)-(A8), leap-frog adopts:
$$\dot{\mathbf{r}}(t+3h/2) = \dot{\mathbf{r}}(t+h/2) + h\ddot{\mathbf{r}}(t+h) \qquad (A9)$$
$$\boldsymbol{\omega}_i(t+3h/2) = \boldsymbol{\omega}_i(t+h/2) + h\dot{\boldsymbol{\omega}}_i(t+h) \qquad (A10)$$
$\dot{\mathbf{r}}(t+h)$, as needed for on-step kinetic energy computation, will be estimated as
$$\dot{\mathbf{r}}(t+h) = \frac{\dot{\mathbf{r}}(t+h/2) + \dot{\mathbf{r}}(t+3h/2)}{2} \qquad (A11)$$
Note that a leap-frog initialisation can be obtained from a velocity-Verlet initial state by means of Eq. (A1) and (A2) (or their variants where $h$ is replaced by $-h$).

## APPENDIX B:
### Thermostatting

The Nosé-Hoover thermostat [65, 66] works by modifying the accelerations in the following way:
$$acceleration\big|_{NVT} = acceleration\big|_{NVE} - \xi \cdot velocity \qquad (B1)$$
NVT stands for constant temperature molecular dynamics and NVE denotes constant energy Newtonian dynamics. $acceleration\big|_{NVE}$ thus stands for the acceleration that would be experienced had the simulation been NVE. $\xi$ is a heat-flow variable with its own equation of motion:
$$\dot{\xi} = (2K - Xk\Theta)/Q. \qquad (B2)$$
$K$ denotes the total kinetic energy of the system and $X$ is the total number of degrees of freedom in the system. $k$ denotes the Boltzmann constant and $\Theta$ is the desired temperature. $Q$ is a parameter with dimension $(energy \cdot mass^2)$ that dictates how rapidly the temperature oscillates about $\Theta$ during the simulation. More specifically, if the time period of that oscillation is desired to be approximately $\tau$, then $Q$ should be set as
$$Q = Xk\Theta\tau^2. \qquad (B3)$$

In the same vein as above, a system of $N$ interacting bola-dimers can be thermostatted by modifying the equations of motion of each dimer as follows:

$$\ddot{\mathbf{r}} = \ddot{\mathbf{r}}|_{NVE} - \xi_P \cdot \dot{\mathbf{r}} \quad (B4)$$

$$\dot{\boldsymbol{\omega}}_i = \dot{\boldsymbol{\omega}}_i|_{NVE} - \xi_i \cdot \boldsymbol{\omega}_i \quad (B5)$$

where $\ddot{\mathbf{r}}|_{NVE}$ and $\dot{\boldsymbol{\omega}}_i|_{NVE}$ denote the $\ddot{\mathbf{r}}$ and $\dot{\boldsymbol{\omega}}$ as computed from Eq. (10) and (11) respectively. Individual dimer indices ($j = 1,...,N$) have been dropped for simplicity, but will be used when needed. Time evolution of the heat-flow variables is governed by:

$$\dot{\xi}_P = \left[ \sum_{j=1}^{N} \left( M\dot{\mathbf{r}}^2 + \dot{\mathbf{r}} \cdot \sum_{i=1}^{2} m_i \boldsymbol{\omega}_i \times \mathbf{r}_i \right)_j - 3Nk\Theta \right] / 3Nk\Theta\tau^2 \quad (B6)$$

and

$$\dot{\xi}_i = \left[ \sum_{j=1}^{N} \left( I_i \boldsymbol{\omega}_i^2 + \dot{\mathbf{r}} \cdot (m_i \boldsymbol{\omega}_i \times \mathbf{r}_i) \right)_j - 2Nk\Theta \right] / 2Nk\Theta\tau^2 \quad (B7)$$

The use of separate heat-flow variables in Eq. (B4) and (B5) above introduces individual temperature controls for the translational and rotational degrees of freedom which helps achieve equilibrium more quickly than would otherwise be possible [66].

Numerical integration of the NVT equations of motion (B4)-(B7) can be readily achieved by modifying a few steps in the velocity-Verlet or the leap-frog algorithm discussed in Appendix A for NVE molecular dynamics. In other words, the NVE integrator can be turned into an NVT one through a few changes only.

Velocity-Verlet scheme for NVT:

*Step* 1: Update velocities to mid-step as:

$$\dot{\mathbf{r}}(t + \frac{h}{2}) = \dot{\mathbf{r}}(t) + \frac{h}{2}\left( \ddot{\mathbf{r}}(t)|_{NVE} - \xi_P(t)\dot{\mathbf{r}}(t) \right) \quad (B8)$$

$$\boldsymbol{\omega}_i(t + \frac{h}{2}) = \boldsymbol{\omega}_i(t) + \frac{h}{2}\left( \dot{\boldsymbol{\omega}}_i(t)|_{NVE} - \xi_i(t)\boldsymbol{\omega}_i(t) \right) \quad (B9)$$

*Step* 2: Same as in Appendix A. In addition, store $\hat{\mathbf{r}}_i(t + \frac{h}{2})$ computed as

$$\hat{\mathbf{r}}(t + \frac{h}{2}) = \frac{\hat{\mathbf{r}}(t+h) + \hat{\mathbf{r}}(t)}{|\hat{\mathbf{r}}(t+h) + \hat{\mathbf{r}}(t)|} \quad (B10)$$

*Step* 3: Same as in Appendix A. Also, compute $\dot{\xi}_P(t + \frac{h}{2})$ and $\dot{\xi}_i(t + \frac{h}{2})$ using Eq. (B6) and (B7) respectively. Finally, propagate the heat-flow variables as:

$$\xi_P(t+h) = \xi_P(t) + h\dot{\xi}_P(t + \frac{h}{2}) \quad (B11)$$

$$\xi_i(t+h) = \xi_i(t) + h\dot{\xi}_i(t + \frac{h}{2}) \quad (B12)$$

*Step* 4: Approximate $\ddot{\mathbf{r}}(t+h)|_{NVE}$ as $\ddot{\mathbf{r}}(t)|_{NVE}$, viz.

$$\ddot{\mathbf{r}}(t+h)|_{NVE} = \ddot{\mathbf{r}}(t)|_{NVE} \quad (B13)$$

and compute $\dot{\boldsymbol{\omega}}_i(t+h)|_{NVE}$ using Eq. (11).

*Step* 5: Estimate $\boldsymbol{\omega}_i(t+h)$ as

$$\boldsymbol{\omega}_i(t+h) = \frac{1}{1 + \frac{h}{2}\xi_i(t+h)}\left( \boldsymbol{\omega}_i(t + \frac{h}{2}) + \frac{h}{2}\dot{\boldsymbol{\omega}}_i(t+h)|_{NVE} \right) \quad (B14)$$

*Step* 6: Compute $\ddot{\mathbf{r}}(t+h)|_{NVE}$ (Eq. (10)) and therefrom determine $\dot{\boldsymbol{\omega}}_i(t+h)|_{NVE}$ (Eq. (11)).

*Step* 7: Repeat step 5.

*Step* 8: Repeat step 6 using the current estimate of $\boldsymbol{\omega}_i(t+h)$ from step 7.

*Step* 9: Update velocities from mid-step to on-step using:

$$\dot{\mathbf{r}}(t+h) = \frac{1}{1 + \frac{h}{2}\xi_P(t+h)}\left( \dot{\mathbf{r}}(t + \frac{h}{2}) + \frac{h}{2}\ddot{\mathbf{r}}(t+h)|_{NVE} \right) \quad (B15)$$

and Eq. (B14).

Molecular dynamics simulations are generally started with the system at rest, i.e. with zero total momentum [64, 69]. The thermostatted equations of motion (B4) and (B5), however, may not preserve total momentum. To keep the system at rest at all times, therefore, it may be necessary to shift all the updated $\dot{\mathbf{r}}$ such that the total momentum remains zero after every time-step.

Leap-frog scheme for NVT:

Leap-frog can be derived from velocity-Verlet as discussed in Appendix A. The final velocity update by a full step (from one mid-step to the next), however, can be done in this case as:

$$\dot{\mathbf{r}}(t + 3h/2) = \frac{\dot{\mathbf{r}}(t+h/2)\left(1 - \frac{h\xi_P(t+h)}{2}\right) + h\ddot{\mathbf{r}}(t+h)|_{NVE}}{\left(1 + \frac{h\xi_P(t+h)}{2}\right)} \quad (B16)$$

$$\boldsymbol{\omega}_i(t + \frac{3h}{2}) = \boldsymbol{\omega}_i(t + \frac{h}{2}) + h\left( \dot{\boldsymbol{\omega}}_i(t+h)|_{NVE} - \xi_i(t+h)\boldsymbol{\omega}_i(t+h) \right) \quad (B17)$$

Note that for $\boldsymbol{\omega}_i(t+h)$ in Eq. (B17) we use its latest estimate (e.g. as determined in step 7 above, which is $O(h^3)$-accurate).

**REFERENCES**

[1] J.H. Fuhrhop, T. Wang, *Chem. Rev.* **104**, 2901-2937 (2004)
[2] A. Meister, A. Blume, *Curr. Opin. Colloid Interface Sci.* **12,** 138–147 (2007)
[3] C.R. Woese, *Sci. Am.* **244**, 94-107 (1981)
[4] A. Gambacorta, A. Gliozzi, M. De Rosa, *World J. Microbiol. Biotechnol.* **11**, 115-131 (1995)
[5] J.L. Gabriel, P.L.G. Chong, *Chem. Phys. Lipids* **105**, 193–200 (2000)
[6] H.C. Jarrell, K.A. Zukotynski, G.D. Sprott, *Biochim. Biophys. Acta* **1369**, 259–266 (1998)
[7] H. Komatsu, P.L.G Chong, Biochem. **37**, 107 –115 (1998)
[8] E.L. Chang, *Biochem Biophys. Res. Comm.* **202**(2), 673-679 (1994)
[9] G.C. Choquet, G.B. Patel, G.D. Sprott. *Can. J. Microbiol.* **42**,183 –186 (1996)
[10] G.D. Sprott et al., *Cells and Materials* **6**(1-3), 143-155 (1996)
[11] G.B. Patel et al., *Int. J. Pharm.* **194**, 39 –49 (2000)
[12] H.J. Freisleben, "Tetraether Lipid Liposomes" (Chapter 8) in *Membrane Structure in Disease and Drug Therapy*, ed. Guido Zimmer, *Marcel Dekker*, NY, 2000
[13] V.P. Chavda, M.M. Soniwala, J.R. Chavda, *Int. J. Pharm. Biol. Sci. Arch.* **1** (1), 38-45 (2013)
[14] G. Réthoré et al., *Chem. Commun.*, 2054–2056 (2007)
[15] L.A. Balakireva, M.Y. Balakirev, "Transfection of Eukaryotic Cells with Bipolar Cationic Derivatives of Tetraether Lipid" (Chapter 9) in *Membrane Structure in Disease and Drug Therapy*, ed. Guido Zimmer, *Marcel Dekker*, NY, 2000
[16] M. Mamusa et al., *Colloids Surf. B: Biointerfaces* **143**, 139–147 (2016)

[17] V. Weissig et al., *J. Liposome Res.* **16**, 249–264 (2006)
[18] G.B. Patel, G.D. Sprott, *Crit. Rev. Biotechnol.* **19**(4), 317–357 (1999)
[19] D.L. Tolson et al., *J. Liposome Res.* **6**(4), 755-776 (1996)
[20] M. Bulacu, X. Périole, S.J. Marrink, *Biomacromolecules* **13**, 196−205 (2012)
[21] B. Schuster et al., *Langmuir* **19**, 2392 –2397 (2003)
[22] M.G.L. Elferink et al., *J. Biol. Chem.* **267**(2), 1375-1381 (1992)
[23] B.A. Cornell et al., *Nature* **387**, 580 –583 (1997)
[24] Y. Kaufman et al., *Langmuir* **29**, 1152−1161 (2013)
[25] M.L. Bode et al., *Chem. Phys. Lipids* **154**, 94–104 (2008)
[26] T. Benvegnu, M. Brard, D. Plusquellec, *Curr. Opin. Colloid Interface Sci.* **8,** 469–479 (2004)
[27] E.A. Runquist, G. M. Helmkamp Jr., *Biochim. Biophys. Acta Biomembranes* **940**, 10–20 (1988).
[28] G. Wang, R.I. Hollingsworth, *J. Org. Chem.* **64**, 4140-4147 (1999)
[29] J.M. Kim, D. H. Thompson, *Langmuir* **8**, 637−644 (1992)
[30] J.M. Delfino, S. L. Schreiber, F.M. Richards, *J. Am. Chem. Soc.* **115**, 3458−3474 (1993)
[31] L.A. Cuccia et al., *Chem. Eur. J.* **6**, 4379-4384 (2000)
[32] S. Svenson, D.H. Thompson, *J. Org. Chem.* **63**, 7180−7182 (1998)
[33] R.A. Moss, T. Fujita, Y. Okumura, *Langmuir* **7**, 2415−2418 (1991)
[34] B. Raguse, et al., *Tetrahedron Lett.* **41**, 2971−2974 (2000)
[35] Y. Yan, T. Lu, J. Huang, *J. Colloid Interface Sci.* **337**, 1–10 (2009).
[36] R. Nagarajan, *Chem. Eng. Comm.* **55**, 251-273 (1987)
[37] F. Cavagnetto et al., *Biochim. Biophys. Acta*, **1106**, 273-28l (1992)
[38] K. Köhler et al., *Angew. Chem. Int. Ed*. **43**, 245 –247 (2004)
[39] A. Meister et al., *J. Phys. Chem. B* **112**, 4506-4511 (2008)
[40] M. Wahab et al., *Langmuir Lett*. **26**(5), 2979–2982 (2010)
[41] A. Meister, A. Blume, *Adv. Planar Lipid Bilayers Liposomes* **16**, 93-128 (2012)
[42] K. Kohler et al., *Soft Matter* **2**, 77–86 (2006)
[43] G. Graf et al., *J. Phys. Chem. B* **115**, 10478–10487 (2011)
[44] A. Meister et al., *Langmuir* **23**, 7715-7723 (2007)
[45] N. Nuraje, H. Bai, K. Su, *Prog. Polym. Sci.* **38** (2), 302-343 (2013)
[46] J.P. Nicolas, *Lipids* **40** (10), 1023-1030 (2005)
[47] W. Shinoda et al., *Biophys. J.* **89**, 3195–3202 (2005)
[48] A.O. Chugunov et al., *Sci. Rep.* 4 : 7462 (2014), DOI: 10.1038/srep07462
[49] Q. Y. Wu, W. Tian, Y. Ma, *J. Phys. Chem. B* **121** (38), 8984-8990 (2017)
[50] S. Li et al., *J. Phys. Chem. B* **113**, 1143–1152 (2009)
[51] S.O. Nielsen et al., *J. Phys.: Cond.. Matt.* 16 R481–R512 (2004)
[52] V. Weissig et al., *Proc. Intl. Symp. Contr. Rel. Bioact. Mat.* **25**, 312 (1998)
[53] R. P. Rand, V. A. Parsegian, *Biochim. Biophys. Acta* **988**, 351 (1989)
[54] I. R. Cooke, M. Deserno, *J. Chem. Phys.* **123**, 224710 (2005)
[55] G. Brannigan, P. F. Phillips, F. L. H. Brown, *Phys. Rev. E* **72**, 011915 (2005)
[56] A. J. Sodt and T. Head-Gordon, *J. Chem. Phys.* **132**, 205103 (2010)
[57] C. Arnarez et al., *J. Chem. Theory Comput.* **11**, 260 (2015)
[58] O. Farago, *J. Chem. Phys.* **119** (1), 596-605 (2003)
[59] J. G. Gay, B. J. Berne, *J. Chem. Phys.* **74**, 3316 (1981)
[60] S. Dey, J. Saha, *Phys. Rev. E* **95**, 023315 (2017)
[61] Such semi-rigid molecules have been discussed by Forester and Smith [T. R. Forester, W. Smith, *J. Comput. Chem.* **19** (1), 102 (1998)]. It may be noted that Bulacu et al. have also made their bolaamphiphiles flexible by relaxing the bond angles associated with a specific bond along the hydrocarbon chain(s) [20].
[62] A. Jain, N. Vaidehi, G. Rodriguez, *J. Comput. Phys.* **106**, 258-268 (1993)
[63] S. Dey, "Time Reversible, Angular Velocity Based Integrator for Rigid Linear Molecules", arXiv:1811.06450 [physics.comp-ph]
[64] M. P. Allen, D. J. Tildesley, *Computer Simulation of Liquids* (Clarendon Press, Oxford, 1987)
[65] W.G.Hoover, *Phys. Rev. A* **31**(3), 1695 (1985)
[66] S.Nosé, *Mol. Phys.* **57**(1), 187 (1986)
[67] L. A. Estroff, A. D. Hamilton**.** *Chem. Rev.* **104**(3), 1201-1217 (2004)
[68] S. Drescher et al., *Langmuir* **30**, 9273−9284 (2014)
[69] D. Frenkel, B. Smit, *Understanding Molecular Simulation* (Academic Press, New York, 1996)
[70] *Coarse-Graining of Condensed Phase and Biomolecular Systems*, edited by G. A. Voth (CRC Press, Taylor & Francis Group, Boca Raton, London, New York, 2009)
[71] D. P. Brownholland et al., *Biophys. J.* **97**(10), 2700-2709 (2009)